%% file: csd.tex
\documentclass[preprint,review]{elsarticle} 
\usepackage[english]{babel}
\usepackage[latin1]{inputenc}

\usepackage{algorithm}
\usepackage{algpseudocode}

\usepackage{graphicx}
 
\usepackage[cmex10]{amsmath}

\usepackage{psfrag}

\usepackage{textcomp}

\usepackage{subfig}
\usepackage{wrapfig}
\usepackage{siunitx} 
\usepackage{float}
\usepackage{multirow}

\usepackage[usenames, dvipsnames]{color} 

\usepackage[english]{varioref}

\usepackage{booktabs}

\usepackage{amsthm} 
\usepackage{gensymb}
\usepackage{amssymb}
\usepackage{fancyhdr}
\usepackage{braket} 

\usepackage{cleveref}

\usepackage{mathtools}

\usepackage{algorithm}
\usepackage{algorithmicx}
\usepackage{algpseudocode}

\usepackage[showframe=false]{geometry}

\usepackage{xcolor,colortbl}

\usepackage{color, soul}

\newtheorem{mycon}{CS Test}
\newtheorem{remark}{Remark}

\newcommand{\TwoRowCell}[1]{ {\scriptsize \begin{tabular}{@{}c@{}} #1 \end{tabular}}  } 

\newcommand{\kx}{\underline{k}}

\algdef{SE}[DOWHILE]{Do}{doWhile}{\algorithmicdo}[1]{\algorithmicwhile\ #1}%

\newcommand{\MAPEpp}{MAPE_{NP}}

\newcommand{\bbm}{\begin{bmatrix}}
\newcommand{\ebm}{\end{bmatrix}}
\newcommand{\etax}{\underline{\eta}}
\newcommand{\etaX}{\overline{\eta}}
\newcommand{\Dalphax}{\underline{\Delta\alpha}}
\newcommand{\DalphaX}{\overline{\Delta\alpha}}
\newcommand{\alphax}{\underline{\alpha}}
\newcommand{\alphaX}{\overline{\alpha}}

\newcommand{\fragCSDGridError}{ %
	\psfrag{ccf}[][][.7][0]{$\beta_0$}
	\psfrag{b0}[][][.7][0]{$\beta_0$}
	\psfrag{MAPEpp}[][][.7][0]{$\MAPEpp$}
	\psfrag{NRMSEpp}[][][.7][0]{$NRMSE_{NP}$}
	\psfrag{MBE}[][][.7]{$MBE$}
	\psfrag{R2}[][][.7]{$R2$}
	}

\newcommand{\PlotLabel}{ \psfrag{kW}{\tiny{kW}}
						 \psfrag{day}{\hspace*{-0.5cm}\tiny{Time (day)}}
						 \psfrag{epoch}{\hspace*{-0.5cm}\tiny{Time (iteration)}}
						 \psfrag{mu1}{\tiny{$\hat{\mu}_1$}}
						 \psfrag{mu2}{\tiny{$\hat{\mu}_2$}}
						 \psfrag{mu3}{\tiny{$\hat{\mu}_3$}}
						 \psfrag{eta2}{\tiny{$\hat{\eta}_2$}}
						 \psfrag{eta3}{\tiny{$\hat{\eta}_3$}}
					   }

\definecolor{Red}{rgb}{1.0,0.6,0.6}
\definecolor{Blue}{rgb}{0.6,0.8,1.0}
\definecolor{Green}{rgb}{0.7,1.0,0.4}
\definecolor{Gray}{gray}{.85}
\newcolumntype{R}{>{\columncolor{Red}}c}
\newcolumntype{b}{>{\columncolor{Blue}}c}
\newcolumntype{g}{>{\columncolor{Green}}c}
\newcolumntype{G}{>{\columncolor{Gray}}c}

\DeclarePairedDelimiter{\abs}{\lvert}{\rvert} 

\journal{Automatica}

\begin{document}
\begin{frontmatter}
\title{Estimation of Photovoltaic Generation Forecasting Models \\ using Limited Information}
\author{Gianni Bianchini}\ead{giannibi@diism.unisi.it} 
\author{Daniele Pepe}\ead{pepe@diism.unisi.it}             
\author{Antonio Vicino}\ead{vicino@diism.unisi.it} 
\address {Dipartimento di Ingegneria dell'Informazione e Scienze Matematiche, Universit\`a di Siena, \\Via Roma 56, 53100 Siena, Italy} 

\begin{abstract}
This work deals with the problem of estimating a photovoltaic generation forecasting model in scenarios where measurements of meteorological variables (i.e. solar irradiance and temperature) at the plant site are not available. A novel algorithm for the estimation of the parameters of the well-known PVUSA model of a photovoltaic plant is proposed. Such a method is characterized by a low computational complexity, and efficiently exploits only power generation measurements, a theoretical clear-sky irradiance model, and temperature forecasts provided by a meteorological service. An extensive experimental validation of the proposed method on real data is also presented.
\end{abstract}

\end{frontmatter}

\section{Introduction}
The electrical grid can be no longer considered a unidirectional
means of distributing energy from
conventional plants to the final users, but a
Smart Grid, where strong interaction between producers
and users takes place \cite{ALBUYEH09}.
A major challenge in the integration of renewable energy sources into the grid \cite{BIB:SCHIFFER2016135} is that power generation is intermittent, difficult to control, and strongly dependent on the variation of weather conditions.
For this reason, forecasting of renewable distributed generation has become a fundamental requirement in order to reliably manage conventional power plant operation,
grid balancing, real-time unit dispatching \cite{BIB:KIM2016193}, demand constraints \cite{BIB:ISHIZAKI2016163}, and energy market requirements. In this respect, renewable generation forecasts on different time horizons are of special interest to various players that operate in the active grid, in particular to Distribution System Operators (DSO) and Transmission System Operators (TSO) (see \cite{ALBUYEH09,BIB:DENHOLM2007_1,BIB:DENHOLM2007_2} and references therein).

Concerning photovoltaic (PV) power generation, researchers have devoted much attention to the problem of obtaining accurate generation forecasts over different time horizons, e.g., day-ahead and hour-ahead \cite{diagne2013review,coimbraoverview}. Most contributions, however, focus on the problem of solar irradiance prediction \cite{Inman2013535,BIB:PEREZ2007,PEREZ2010,BIB:KANG2015}. To tackle this problem, several approaches based on Artificial Neural Networks (ANNs) \cite{CAPIZZI12,WU11,BIB:CORNARO} or Support Vector Machines \cite{RAGNACCI12} can be found in the literature. 
Alternatively, classical linear time series prediction methods are used in \cite{REIKARD09,LORENZ09,BACHER09}, where the considered time series is typically the global horizontal irradiance (GHI) \cite{WONG01,ASHRAE09}. GHI forecasts are typically used along with temperature forecasts in a simulation model of the PV plant \cite{PATEL06} in order to calculate generated power predictions. In all cases, computing reliable forecasts from predicted meteorological variables hinges upon the availability of an accurate model of the plant, be it physical or estimated from data.

Unfortunately, in many common scenarios, neither a plant model, nor direct on-site measurements of solar irradiance and other meteorological variables (e.g., temperature) are available. This is always the case with a DSO dealing with hundreds or thousands of heterogeneous, independently owned and operated PV plants; in this case, the only available data consists of generated power measurements provided by electronic meters, and of irradiance and temperature forecasts provided by a meteorological service. The problem of forecasting power generation in this case is addressed in \cite{SHANXU10} by means of a neural network and in \cite{pepe2017model,BIB:ENERGYCON16} using a parametric model. In these approaches, however, further information on the cloud cover index  at the plant site is assumed to be available. In \cite{BIB:ISGT,BIB:CDC13}, a heuristic method for the estimation of the parameters of well-known PVUSA model \cite{Dows95} based on theoretical clear-sky irradiance is presented, while in \cite{BIB:ENERGYCON18}, a recursive procedure based on the clear-sky criteria proposed in \cite{BIB:RENO2016520} is devised. However, the former approach does not allow for capturing possible parameter variations or seasonal drifts, and moreover both approaches require trial-and-error in order to manually tune a number of algorithm parameters whose values may vary significantly according to the climate zone.

\subsection{Paper contribution}
In this paper, a novel approach to the problem of estimating the parameters of the PVUSA model in the partial information case is presented. The only historical data used by the method consist of generated power, and temperature (but not irradiance) forecasts. Parameter estimates are obtained by carefully exploiting the information contained in portions of the generated power data which turn out to be meaningful if combined with theoretical clear-sky irradiance over the same period. More specifically, we introduce three tests to be  performed on generated power data in order to detect portions of such data that were generated under clear-sky conditions. The information contained in such portions is then exploited in a recursive parameter estimation algorithm in combination with theoretical clear-sky irradiance provided by a suitable model. The method proposed in this paper improves over \cite{BIB:ISGT,BIB:CDC13,BIB:ENERGYCON18}, since it is able to adapt to parameter variations and requires the tuning of a single threshold coefficient whose physical role can be interpreted in terms of the cloud cover factor (CCF) \cite{BIB:KASTEN}.

The paper is structured as follows: in Section \ref{PRE} the modeling tools are introduced; in Section \ref{SEC:CS_DETECTION} the proposed clear-sky detection tests are developed; the model estimation procedure is presented in Section \ref{sec:estimation}. In Section~\ref{sec:forecasting}, the relevant forecasting problems are recalled, and performance evaluation criteria are discussed in Section~\ref{SEC:PEVA}. Experimental validation results are reported in Section~\ref{SEC:EXP_RES}, and conclusions are drawn in Section~\ref{sec:CONC}.

\section{Models and methods}\label{PRE}
\subsection{The PVUSA photovoltaic plant model}
\label{SEC:PVUSA}
A PV plant can be efficiently modelled using the PVUSA model~\cite{BIB:PVUSA}, which expresses the instantaneous generated power as a function of irradiance and air temperature according to the equation:
\begin{equation}
\label{EQ:PVUSA}
P = \mu_1 I + \mu_2 I^2 + \mu_3 IT,
\end{equation}
where $P$, $I$, and $T$ are the generated power (\si{\kilo\watt}), irradiance (\si{\watt/\metre^{2}}), and air temperature (\si{\degreeCelsius}), respectively, and $\mu_1$, $\mu_2$, $\mu_3$, are the model parameters. It is important to notice that
model \eqref{EQ:PVUSA} is linear in the parameters. For the purpose of this work, it is useful to express \eqref{EQ:PVUSA} in the form
\begin{equation}
\label{EQ:PVUSAK}
P = \mu_1 \cdot \alpha(I,T) \cdot I,
\end{equation}
where
\begin{equation}
\label{EQ:ALPHA}
\alpha(I,T) = 1+ \eta_2 I + \eta_3 T,
\end{equation}
being
\begin{equation}\label{EQ:ETA}
\eta_2=\mu_2/\mu_1,\;\; \eta_3= \mu_3/\mu_1.
\end{equation}
From \eqref{EQ:PVUSAK}, it is apparent that $\mu_1$ represents the main power/irradiance gain of the plant, while $\alpha(I,T)$ in \eqref{EQ:ALPHA} can be seen as correction term. In this respect, it is worth noticing that the ratios $\eta_2$  and $\eta_3$  in \eqref{EQ:ETA} are characterized by well-established variability ranges among different PV technologies (see \cite{BIB:PVUSA}). Such ranges are given by:
\begin{equation}
\begin{split}
\eta_2 &\in \left[ \underline \eta_2, \overline \eta_2\right]  = \left[\num{-2.5e-4}, \num{-1.9e-5}\right], \\
\eta_3 &\in \left[ \underline \eta_3, \overline \eta_3\right]  = \left[\num{-4.8e-3}, \num{-1.7e-3}\right].
\end{split}
\label{EQ:AtoBC}
\end{equation}
This property will be exploited in the proposed estimation procedure. It is also appropriate to represent \eqref{EQ:PVUSA} also in the standard regressive form
\begin{equation}
\label{EQ:PVUSAKREG}
P = \phi'(I,T)~\mu,
\end{equation}
where the regressor is given by
\begin{equation}\label{eq:regressor}
\phi(I,T)=[I~~I^2~~IT]',
\end{equation}
and the parameter vector is
\begin{equation}\label{eq:params}
\mu=[\mu_1~~\mu_2~~\mu_3]'.
\end{equation}

The PVUSA model can be fruitfully exploited for the purpose of computing forecasts of generated power on the basis of predicted meteorological variables. Indeed, once a correct estimate $\hat\mu$ of the parameter vector is available, a reliable power generation forecast $\hat P$ can be obtained by substituting predicted irradiance $\hat I$ and temperature $\hat T$, provided by a meteorological service, into the model equation \eqref{EQ:PVUSAKREG}, i.e.,
\begin{equation}
\label{EQ:PVUSAKREGF}
\hat P = \phi'(\hat I,\hat T) \!\cdot \hat\mu.
\end{equation}

Similarly, generation forecasts under clear-sky conditions can be obtained by using the theoretical irradiance $I^{cs}$ at the plant location, as provided by a suitable model, and temperature forecasts, i.e.,
\begin{equation}\label{eq:cspredX}
\hat P^{cs} = \phi'(I^{cs},\hat T) \!\cdot \hat\mu.
\end{equation}
Notice that clear-sky generation forecasts provide an upper bound on the power that can be generated by a plant, and as such they can be used by the DSO, for instance, when scheduling the maintenance of the portion of the grid where the plant is located.
Despite its simplicity, very good forecasting accuracy is obtained from the PVUSA model when the parameter vector $\mu$ is estimated using measured irradiance and temperature data via, e.g., standard least squares fitting (see, e.g.,~\cite{BIB:ISGT}).


A DSO that manages a high number of independent generation facilities may not have access to time series of irradiance and temperature measured on the premises of each plant, while power generation data are always available through meters. In order to estimate model parameters, replacing the measured values of $I$ and $T$ with forecasts $\hat{I}$ and $\hat{T}$ provided by a meteorological service is not a viable solution, due to the fact that forecasting errors on the irradiance are in general too large. On the contrary, temperature forecasts are quite reliable and can be used in place of actual measurements (see \cite{BIB:ISGT,BIB:CDC13} for details).


\subsection{Clear-sky irradiance model}
In this paper, a theoretical estimate of the global clear-sky irradiance on a given surface is required. To this aim, although several different models are present in the literature \cite{ineichen06}, the Heliodon simulator model \cite{BIB:HELIODON} is used. This model is characterized by a high degree of simplicity and allows to compute the theoretical global clear-sky normal irradiance (\si{\watt/\meter^2}) from the solar altitude~$h$, i.e., the angle over the horizon (rads), as:
\begin{equation}
\label{EQ:ELIODON}
I^{cs,n} = 
\begin{cases}
A\cdot0.7^{ \left( \frac{1}{\sin{h}} \right)^{0.678}} 	& \text{if } 0<h<\pi/2 \\
0																			& \text{otherwise,}
\end{cases}	
\end{equation}
where $A=1353$ \si{\watt/\meter^2} denotes the apparent extraterrestrial irradiance. Given the theoretical clear-sky normal irradiance $I^{cs,n}$, the clear-sky irradiance on an inclined panel surface $I^{cs}$ can be derived from $I^{cs,n}$ and the orientation of the surface with respect to the sun position. Denoting by $\zeta$ the surface azimuth and $\psi$ the surface tilt angle, one has that
\begin{equation}
\label{EQ:ICS}
I^{cs}=\left[\sin(\psi)\cos(h)\cos(\zeta-\gamma)+\cos(\psi)\sin(h)\right] I^{cs,n},
\end{equation}
where $\gamma$ is the solar azimuth. Clearly, $I^{cs}$ can be computed for given values of $\zeta$ and $\psi$ from latitude, longitude and time of day. For $\psi=0$, the irradiance on a horizontal surface is obtained.

In this study, it is assumed that the exact orientation of the PV panel surfaces of the considered plant is not known a-priori. However, it is reasonable to suppose that the plant is efficiently oriented for the specific latitude according to, e.g., the guidelines given in~\cite{BIB:ORIENTATION}. Therefore, the value of (\ref{EQ:ICS}) with $(\zeta,\psi)$ taken from the above guidelines will be used as a reference for the theoretical clear-sky irradiance $I^{cs}$ for a given plant.

%
%
%

\section{Clear-sky data detection}
\label{SEC:CS_DETECTION}
In this paper, the following key idea is exploited for the purpose of estimating the parameters of the PVUSA model \eqref{EQ:PVUSA} of a PV plant without resorting to on-site irradiance measurements. Given a time series composed of generated power measurements and temperature forecasts (or measurements, if available), suitable tests can be performed on the data in order to detect portions of the power curve which have been generated under a clear-sky condition; this allows for fitting the parameters of the PVUSA model to such data by using theoretical clear-sky irradiance (e.g., via the model \eqref{EQ:ELIODON},\eqref{EQ:ICS}) in the regressor of \eqref{eq:regressor} in place of the actual measured irradiance. This section deals with the derivation of such tests, which will be referred to as {\sl CS tests} in the sequel.

In view of \eqref{EQ:AtoBC}, suitable bounds can be derived on $\alpha(I,T)$ and $P$ in the PVUSA model \eqref{EQ:PVUSAK}-\eqref{EQ:ALPHA}. Indeed, from  \eqref{EQ:ALPHA} and \eqref{EQ:AtoBC}, it is easily checked  that 
\begin{equation}
\underline{\alpha}(I,T) \leq \alpha(I,T) \leq \overline{\alpha}(I,T),
\label{eq:alpha_bound}
\end{equation}
where
\begin{align}
\underline{\alpha}(I,T) &= 
\begin{cases}
1 + \underline{\eta}_2I + \underline{\eta}_3T, & \text{if $T \geq 0$} \\
1 + \underline{\eta}_2I + \overline{\eta}_3T, & \text{if $T < 0$}
\end{cases}\label{eq:alpha_min}\\
\overline{\alpha}(I,T) &= 
\begin{cases}
1 + \overline{\eta}_2I + \overline{\eta}_3T, & \text{if $T \geq 0$} \\
1 + \overline{\eta}_2I + \underline{\eta}_3T, & \text{if $T < 0$.}
\end{cases}\label{eq:alpha_max}
\end{align}
Moreover, for realistic values of $I$ and $T$, it always holds that $\underline{\alpha}(I,T)>0$ and $\overline{\alpha}(I,T)<1$. From \eqref{eq:alpha_bound} and \eqref{EQ:PVUSAK}, the following bound on $P$ is obtained: 
\begin{equation}
\label{eq:pwr_bound}
\mu_1 \cdot I \cdot \underline{\alpha}(I,T) \leq P \leq \mu_1 \cdot I \cdot \overline{\alpha}(I,T).
\end{equation}

Let us now consider a time series $\{P(j)$, $I(j)$, $T(j)\}$ of the variables in \eqref{EQ:PVUSA}, where $j$ represents a discrete time index. The increment of $P(j)$ can be expressed as
\begin{equation}\label{eq:deltaP}
\begin{split}
\Delta P(j) & = P(j) - P(j-1) \\
& = \mu_1 \left[ I(j-1)\Delta \alpha(j) + \Delta I(j)\alpha(I(j),T(j)) \right],
\end{split}
\end{equation}
where
\begin{align*}
&\Delta I(j) = I(j) - I(j-1), \\
&\Delta \alpha(j) = \alpha(I(j),T(j)) - \alpha(I(j-1),T(j-1)).
\end{align*}

Let $\Delta T(j) = T(j) - T(j-1)$. Taking into account \eqref{eq:alpha_min}-\eqref{eq:alpha_max}, it is easily checked that the following bounds on $\Delta \alpha(j)$ hold:
\begin{equation}
\label{eq:bound_delta_alpha}
\underline{\Delta\alpha}(j) \leq \Delta \alpha(j) \leq \overline{\Delta\alpha}(j),
\end{equation}
where 
\begin{equation}
\left[\begin{array}{c} 	\underline{\Delta\alpha}(j) \\ 	\overline{\Delta\alpha}(j) \end{array}\right] = Q(j) \left[\begin{array}{c} 	\Delta I(j) \\ 	\Delta T(j) \end{array}\right] 
\end{equation}
and the matrix $Q(j)$ depends on the signs of $\Delta I(j)$ and $\Delta T(j)$ according to the following table:
\begin{equation}\notag
\begin{array}{c|c|c}
\toprule
Q(j) & \Delta T(j) \geq 0 & \Delta T(j) < 0 \\
\midrule
\Delta I(j) \geq 0 & \bbm \etax_2 & \etax_3 \\ \etaX_2 & \etaX_3 \ebm & \bbm \etax_2 & \etaX_3 \\ \etaX_2 & \etax_3  \ebm \\
\midrule
\Delta I(j) < 0 & \bbm \etaX_2 & \etaX_3 \\ \etax_2 & \etax_3 \ebm & \bbm \etaX_2 & \etaX_3 \\ \etax_2 & \etaX_3 \ebm \\
\bottomrule	
\end{array}	
\end{equation}
In view of \eqref{eq:deltaP}, this allows to derive the following bounds on $\Delta P(j)$:
\begin{align}
\mu_1\underline{\delta}_{P}(j) &\leq \Delta P(j) \leq \mu_1\overline{\delta}_{P}(j), \label{eq:grad_pwr}
\end{align}
where
\begin{equation}\label{eq:deltapi}
\left[\begin{array}{c} \underline{\delta}_{P}(j) \\ 	\overline{\delta}_{P}(j) \end{array}\right] = R(j) \left[\begin{array}{c} 	I(j-1) \\ 	\Delta I(j) \end{array}\right] 
\end{equation}
and the matrix $R(j)$, depending on the sign of $\Delta I(j)$, is given by 
\begin{equation}\notag
\begin{array}{c|c}
\toprule
R(j) & ~ \\
\midrule
\Delta I(j) \geq 0	& \bbm \Dalphax(j) & \alphax\left(I(j),T(j)\right) \\ \DalphaX(j) & \alphaX\left(I(j),T(j)\right) \ebm \\
\midrule
\Delta I(j) < 0		& \bbm \Dalphax(j) & \alphaX\left(I(j),T(j)\right) \\ \DalphaX(j) & \alphax\left(I(j),T(j)\right) \ebm \\
\bottomrule
\end{array}
\end{equation}

The bounds \eqref{eq:pwr_bound} and \eqref{eq:grad_pwr} allow to devise the sought CS tests. 
Let us consider a time interval $\mathcal J$, and the following associated time series 
\begin{equation}\label{eq:data}
{\mathcal P}_{\mathcal J} =\left\{ \{P^m(j),T(j),P^{cs}(j) \right\},\;j\in{\mathcal J}\}, 
\end{equation}
where, for each time instant $j$, $P^m(j)$ represents the measured plant power reported by meters, $T(j)$ is a temperature forecast (or measurement), and $P^{cs}(j)$ is the clear-sky generated power predicted by a PVUSA model characterized by given values of the parameters $\mu_1,\mu_2$, and $\mu_3$, i.e., 
\begin{align}
P^{cs}(j) &= \mu_1 I^{cs}(j) \alpha\left( I^{cs}(j), T(j) \right),
\end{align}
where the clear-sky irradiance $I^{cs}(j)$ is computed, e.g., via \eqref{EQ:ELIODON}.
Clearly, by \eqref{eq:pwr_bound},
\begin{equation}\begin{split}
\label{eq:cs_bound}
\mu_1 \cdot I^{cs}(j) \cdot \underline{\alpha}\left(I^{cs}(j),T(j)\right) \leq P^{cs}(j) \\
\leq \mu_1 \cdot I^{cs}(j) \cdot \overline{\alpha}\left(I^{cs}(j),T(j)\right).
\end{split}
\end{equation}
Now let
\begin{align}
j_{max} &= \arg \max_{j \in \mathcal{J}} \lbrace I^{cs}(j)  \rbrace, \\
I^{cs}_{max} &= I^{cs}(j_{max}), \\
P^{cs}_{max} &=  P^{cs}(j_{max}) = \mu_1 I^{cs}_{max}\alpha\left(I^{cs}_{max},T(j_{max})\right).
\end{align}
The quantities $I^{cs}_{max}$, $P^{cs}_{max}$, and $j_{max}$ define, respectively, the maximum clear-sky irradiance, the maximum predicted clear-sky generated power, and the time index for which this maximum value occurs within the given time window $\mathcal{J}$.\\
Normalizing~\eqref{eq:cs_bound}  with respect to $P^{cs}_{max}$ yields
\begin{equation}
\begin{split}
\label{eq:cs_norm_bound}
\frac{ I^{cs}(j) \cdot \alphax\left(I^{cs}(j),T(j)\right)}{ I^{cs}_{max} \cdot \alpha(I^{cs}_{max},T(j_{max})) } \leq
\frac{ P^{cs}(j) }{ P^{cs}_{max} } \\ 
\leq
\frac{ I^{cs}(j) \cdot \alphaX\left(I^{cs}(j),T(j)\right) }{ I^{cs}_{max} \cdot \alpha\left(I^{cs}_{max},T(j_{max})\right) },
\end{split}
\end{equation}
and hence the following bounds on the ratio $\frac{ P^{cs}(j) }{ P^{cs}_{max} }$ hold:
\begin{equation}\label{eq:cs_norm_bound2}
\underline{\gamma}_1(j)  \leq
\frac{ P^{cs}(j) }{ P^{cs}_{max} } \leq
\overline{\gamma}_1(j) ,
\end{equation}
where
\begin{equation}\label{eq:gammas}
\begin{split}
\underline{\gamma}_1(j) &= \frac{ \underline{\alpha}( I^{cs}(j), T(j) ) }{ \overline{\alpha}( I^{cs}_{max}, T(j_{max}) ) } \cdot \frac{  I^{cs}(j) }{  I^{cs}_{max} }, \\
\overline{\gamma}_1(j) &= \frac{ \overline{\alpha}( I^{cs}(j), T(j) ) }{ \underline{\alpha}( I^{cs}_{max}, T(j_{max}) ) } \cdot \frac{  I^{cs}(j) }{  I^{cs}_{max} }.
\end{split}
\end{equation}
It is important to observe that \eqref{eq:cs_norm_bound2}-\eqref{eq:gammas} define bounds on the clear-sky power time series which do not depend on the model parameters.
Condition \eqref{eq:cs_norm_bound2} can be exploited in order to classify a time window $\mathcal J$ of measured power data points as generated under clear-sky conditions. Indeed, given the time series $\{P^m(j)$,$T(j)$, $j\in{\mathcal J}\}$, the following test is introduced:
\begin{mycon}\label{con:uno}
	\begin{equation}\label{eq:test1}
	\underline{\gamma}_1 (j) \leq
	\frac{ P^{m}(j) }{ P^{m}(j_{max}) } \leq
	\overline{\gamma}_1(j), \quad \forall j\in\mathcal{J} .
	\end{equation}
\end{mycon}

The satisfaction of CS test 1 is in general not sufficient to classify power data within~${\mathcal J}$ as having been generated under a clear-sky condition. Specifically, if the sky is partially cloudy during the time interval~${\mathcal J}$, the measured power may heavily oscillate, but could remain quite close to the clear-sky power at the maximum~\cite{BIB:RENO2016520}, thus satisfying \eqref{eq:test1}.
To overcome this issue, a further condition on the normalized increment of the  power time series is derived. Let $\underline{\delta}_{P}^{cs}(j)$ and $\overline{\delta}_{P}^{cs}(j)$ be defined by \eqref{eq:deltapi} evaluated for $I(j)=I^{cs}(j)$ and $\Delta I(j)=\Delta I^{cs}(j)= I^{cs}(j) - I^{cs}(j-1)$. The increment of $P^{cs}(j)$ is given by
\[
\Delta P^{cs}(j) = P^{cs}(j) - P^{cs}(j-1)
\]
and satisfies
\begin{align}
\mu_1\underline{\delta}_{P}^{cs}(j) &\leq \Delta P^{cs}(j) \leq \mu_1\overline{\delta}_{P}^{cs}(j) \label{eq:grad_pwrcs}
\end{align}
by \eqref{eq:grad_pwr}.
Normalizing ~\eqref{eq:grad_pwrcs} with respect to $P^{cs}_{max}$, the following bounds on the normalized increment $\frac{ \Delta P^{cs}(j) }{ P^{cs}_{max} }$ are obtained:
\begin{equation}
\label{eq:cs_norm_grad}
\frac{  \underline{\delta}_{P}^{cs}(j) }{ I^{cs}_{max}\overline{\alpha}(j_{max}) } \leq
\frac{ \Delta P^{cs}(j) }{ P^{cs}_{max} } \leq
\frac{  \overline{\delta}_{P}^{cs}(j) }{  I^{cs}_{max} \underline{\alpha}(j_{max}) },
\end{equation}
i.e.,
\begin{equation}
\label{eq:cs_norm_grad_b}
\underline{\gamma}_2 (j) \leq
\frac{ \Delta P^{cs}(j) }{ P^{cs}_{max} } \leq
\overline{\gamma}_2 (j)
\end{equation}
where
\begin{equation}\label{eq:gamma2s}
\begin{split}
\underline{\gamma}_2 (j) & = \frac{ \underline{\delta}_{P}^{cs}(j) }{ \overline{\alpha}(j_{max}) } \cdot \frac{ 1 }{ I^{cs}_{max}} \\
\overline{\gamma}_2 (j) & = \frac{ \overline{\delta}_{P}^{cs}(j) }{ \underline{\alpha}(j_{max}) }  \cdot \frac{ 1 }{ I^{cs}_{max}} .
\end{split}
\end{equation}
Note that the bounds \eqref{eq:cs_norm_grad_b}-\eqref{eq:gamma2s}, similarly to \eqref{eq:cs_norm_bound2}-\eqref{eq:gammas}, do not depend on the model parameters.
Condition \eqref{eq:cs_norm_grad_b} provides the second criterion for classifying a time window ${\mathcal J}$ of measured power data points as clear-sky. The following test is introduced:
\begin{mycon}\label{con:due}
	\begin{equation}
	\label{eq:cs_norm_grad}
	\underline{\gamma}_2 (j) \leq
	\frac{ \Delta P^{m}(j) }{ P^{m}(j_{max}) } \leq
	\overline{\gamma}_2 (j), \quad \forall j\in{\mathcal J} ,
	\end{equation}
\end{mycon}
\noindent where $\Delta P^m(j)$ is the increment of the measured power, i.e., $\Delta P^m(j) = P^m(j) - P^m(j-1)$.

CS tests~1 and~2 detect deviations in the shape of the normalized power curve from the clear-sky condition caused by cloudiness in different scenarios. However, due to normalization, such conditions may turn out to be fulfilled on a given time window ${\mathcal J}$ when the corresponding data are generated under perfectly uniform cloudiness, i.e., when the actual irradiance satisfies 
\begin{equation}\label{eq:uniform}
I(j) = \beta I^{cs}(j) \quad \forall j \in \mathcal{J},
\end{equation}
where $0<\beta<1$ is a constant that represents a uniform cloud cover factor (see~\cite{BIB:KIMURA}) in the time window $\mathcal{J}$.
If the data collected within such a time window are used to perform a model parameter adaptation step in a recursive estimation procedure, the algorithm may tend to underestimate the power/irradiance gain of the plant at such step. This fact may be detrimental when a long series of data collected under uniform cloudiness is processed. To mitigate this effect, a further test is introduced. Suppose that a current estimate $\hat \mu$ of the model parameters is available. Accordingly, a current estimate of the generated power under clear-sky conditions is given by
\[
\hat P^{cs}(j) = \phi'\left(I^{cs}(j), T(j)\right) \cdot \hat\mu = \hat\mu_1 \cdot I^{cs}(j) \cdot \hat\alpha\left(I^{cs}(j),T(j)\right).
\]
Let $\hat P^{cs}_{max}$ be the peak value of $\hat P^{cs}(j)$ in $\mathcal J$, i.e.,
\[
\hat P^{cs}_{max} = \hat P^{cs}(j_{max}).
\]
Provided that CS tests 1 and 2 are passed by the data in time window $\mathcal J$, the following further condition is introduced, which involves a comparison of the maximum currently predicted clear-sky power $\hat P^{cs}_{max}$ with the corresponding generated power $P^m(j_{max})$ as follows: 
\begin{mycon}\label{con:tre}
	\begin{equation}
	\label{eq:pmax}
	\frac{P^m(j_{max})}{\hat P^{cs}_{max}} \geq {1-\epsilon},
	\end{equation}
\end{mycon}
\noindent where $0<\epsilon<1$ is a parameter chosen by the designer, typically a number slightly higher than 0, whose role and choice is discussed later on. CS test \ref{con:tre} has the specific role of detecting, under the condition that CS tests 1 and 2 are satisfied, whether the peak value of measured power within the considered time window lies above a given fraction of the clear-sky power currently estimated by the model. Condition \eqref{eq:pmax} can be satisfied in the following cases:
\begin{itemize}
	\item $\frac{P^m(j_{max})}{\hat P^{cs}_{max}} \geq 1$: in this case the model is currently underestimating clear-sky power;
	\item $ 1-\epsilon \leq \frac{P^m(j_{max})}{\hat P^{cs}_{max}} < 1$ and the current model is overestimating the generated power by a small amount;
	\item $ 1-\epsilon \leq \frac{P^m(j_{max})}{\hat P^{cs}_{max}} < 1$ and uniform cloudiness is present within the given time window, so that generation is marginally lower than the clear-sky power currently predicted by the model. 
\end{itemize}
With the exception of the latter case, the simultaneous satisfaction of CS tests 1,2, and 3 requires that the model parameters be adapted in order to fit the measured power series with the predicted one within $\mathcal J$.

\begin{remark}\label{rem:params}{\rm
		The parameter $\epsilon$ plays a key role in detecting whether the clear-sky curve provided by the model matches or underestimates power data satisfying CS tests 1 and 2, which are related to the shape of the normalized power curve. Setting this value very close to zero allows for good adaptation when the model is underestimating the clear-sky power (for this reason it is advisable to choose an underestimate of $\mu_1$ as the initial guess in the estimation procedure, as detailed in the next section). Higher values, on the other hand, allow for adjusting the model when it is overestimating; the latter case is very important for capturing possible slow  parameter drifts as well as seasonal variations in the accuracy of the theoretical clear-sky model. However, increasing $\epsilon$ may cause adaptation to long series of data generated under uniform cloudiness.
		To further clarify this aspect, let us assume that the true plant is described by a PVUSA model \eqref{EQ:PVUSAK} characterized by $\mu_1=\mu_1^0$ and $\alpha(I,T)=\alpha^0(I,T)$, and that uniform cloudiness is present within $\mathcal J$ so that \eqref{eq:uniform} holds for some $0<\beta<1$. It follows that
		\[
		P^{m}(j_{max}) = \mu_1^0 \cdot \beta \cdot I^{cs}_{max} \cdot \alpha^0\left( I^{cs}_{max}\beta, T(j_{max} \right) ).
		\]
		Therefore condition~\eqref{eq:pmax} becomes:
		\begin{equation}\label{eq:stimaccf}
		\beta \cdot \frac{\mu_1^0}{\hat{\mu}_1} \cdot \frac{ \alpha^0\left( I^{cs}_{max}\beta, T(j_{max} ) \right) }{ \hat \alpha\left( I^{cs}_{max}, T(j_{max} ) \right) } \geq 1 - \epsilon.
		\end{equation}
		For given $\epsilon$, a rough estimate of the values of the uniform cloud cover factor $\beta$ for which CS test 3 is satisfied can be obtained by observing that $\frac{ \alpha^0( I^{cs}_{max}\beta, T(j_{max} ) ) }{ \hat \alpha( I^{cs}_{max}, T(j_{max} ) ) }\approx 1$ (especially for $\beta$ close to 1) and that a rough approximation of the main power/irradiance gain $\mu_1^0$ is given by the ratio $P_{nom}/1000$, where $P_{nom}$ denotes the nominal plant power in kW. In view of \eqref{eq:stimaccf}, CS test 3 is passed when $\beta$ approximately satisfies
		\begin{equation}\label{eq:stimaccf2}
		\beta\gtrapprox \hat\mu_1 \cdot \frac{1000}{P_{nom}}\cdot\left(1-\epsilon\right).
		\end{equation}
		The relationship \eqref{eq:stimaccf2} provides an interpretation of the parameter $\epsilon$ and represents a possible guideline for tuning such parameter on the basis of the minimum value of the cloud cover factor for which the designer allows data generated under uniform cloudiness to be considered for parameter adaptation. However, the effect of the choice of $\epsilon$ will be extensively discussed on the basis of real data in the experimental section of this paper. 
	}
\end{remark}

\section{Model estimation}\label{sec:estimation}

According to the observations in the previous sections, we now introduce the proposed PVUSA plant model estimation method, which yields an on-line update of the parameter vector estimate $\hat\mu$ by relying only on the information contained on a time series composed by measured power $P^m$ and forecast (or measured) temperature $T$.  The model estimation procedure is recursive, and combines CS tests $1-3$ with a standard Recursive Least-Squares (RLS) algorithm using a dynamical time window.

The following definitions are instrumental for building up the procedure:
\begin{itemize}
	\item $k$: present time index;
	\item $d$: present day;
	\item ${\mathcal I}_d=[\underline k_d,\overline k_d]$: time interval corresponding to light hours in day $d$, i.e., $I^{cs}(k)>0$ for all $\underline k_d\leq k \leq \overline k_d$;
	\item ${\mathcal J}_{k,l}$: set of time indices corresponding to a time window of given length $l$ ending at $k$, i.e., ${\mathcal J}_{k,l}=\{k-l+1,\dots,k\}$;
	\item $\hat\mu(k)$: estimate of the parameter vector at time $k$, being $\hat\mu(0)$ the initial guess;
	\item $I^{cs}(j)$: theoretical clear-sky solar irradiance at time step $j$, computed according to \eqref{EQ:ELIODON},\eqref{EQ:ICS}, or a different model for the plant site;
	\item ${T}(j)$: temperature forecast (or measurement, if available) at time $j$ at the plant site, provided by a meteorological service;
	\item $P^m(j)$: measured generated power at time $j$;
	\item $D(j) = \{P^m(j),T(j),I^{cs}(j)\}$: data sample at time $j$;
	\item ${\mathcal D}({\mathcal J})=\left\{ D(j),~ j\in{\mathcal J} \right\}$: data set pertaining to time window ${\mathcal J}$;
	\item $\epsilon$: fixed threshold value $(0<\epsilon<1$);
	\item $l_{min}$: minimum time window length.
	%
\end{itemize}

\begin{figure*}[!htb]
	\centering
	\includegraphics[]{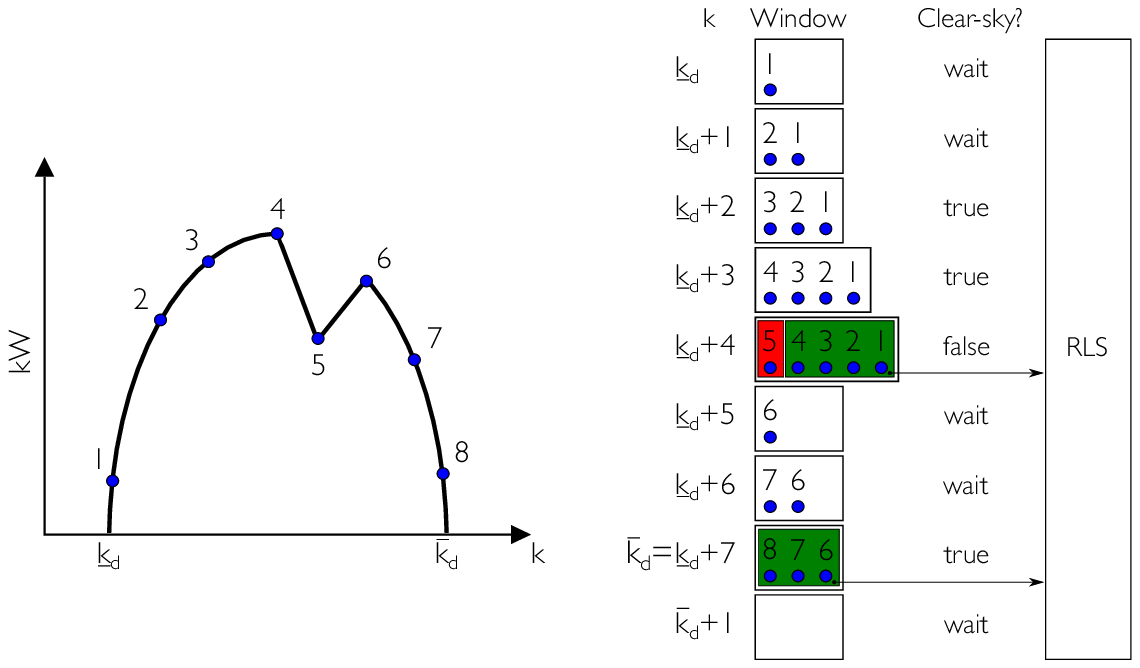}
	\caption{Visual representation of Algorithm~\ref{algorithm}. On the left, measured power data are sketched. On the right, the construction of the data set ${\mathcal D}({\mathcal J}_{k,l})$ is shown. Note that ${\mathcal D}({\mathcal J}_{\underline k_d+4,5})$ is not classified as clear-sky data, while ${\mathcal D}({\mathcal J}_{\underline k_d+3,4})$ is, and parameters are adapted using the latter. At $k=\underline k_d+7$ the day ends. In this particular case $l_{min}=3$.}	
	\label{fig:algorithm}
\end{figure*}

The estimation algorithm is constructed as follows (see Figure \ref{fig:algorithm}). The procedure is reset on each day $d$ at time $k=\underline k_d$. The current parameter estimate $\hat\mu(\underline k_d)$ is initialized with the last estimate obtained on day $d-1$. An initial data set ${\mathcal D}({\mathcal J}_{k,l_{min}})$ is constructed at time $k=\underline k_d +l_{min} -1$ corresponding to an initial time window ${\mathcal J}_{k,l_{min}}$ of length $l_{min}$. If ${\mathcal D}({\mathcal J}_{k,l_{min}})$ does not pass CS tests $1-3$, then the procedure is reset at time $k=\underline k_d + 1$. Otherwise (i.e., if ${\mathcal D}({\mathcal J}_{k,l_{min}})$ is recognized as generated under clear-sky), a new data sample $D(k)$ is acquired at each following step $k$ and added to the current data set ${\mathcal D}({\mathcal J}_{k,l})$, incrementing the length of the time window ${\mathcal J}_{k,l}$ by one. Then, CS tests $1-3$ are performed on ${\mathcal D}({\mathcal J}_{k,l})$. If tests are passed, then further data samples are added to the data set until one of the tests fails (or the end of the day is reached) at some time $k'$. When this occurs, the data set ${\mathcal D}({\mathcal J}_{k'-1,l-1})$ is deemed to be generated under clear-sky conditions and an RLS adaptation step is performed using such data in order to obtain an updated parameter estimate $\hat\mu(k')$. Then, the algorithm is reset at time $k=k'$ and repeated. A detailed description of the procedure is reported in Algorithm \ref{algorithm}.
\begin{algorithm}
	\caption{Parameter estimation}\label{algorithm}
	\begin{algorithmic}[1]
		\small
		\State On each day $d$
		\State $k' \leftarrow \kx_d$
		\While{$k'+l_{min}-1 \leq k_d$}
		\For{$k=k':k'+l_{min}-1$} \Comment{\textit{Get the initial data set} ${\mathcal D}({\mathcal J}_{k,l_{min}})$}
		\State Acquire $D(k)$
		\EndFor
		\If{${\mathcal D}({\mathcal J}_{k,l_{min}})$ does not satisfy CS 1-3} 
		\State $k'\leftarrow k' + 1$ \Comment{$D(k')$ \textit{is rejected and the algorithm is reset at time} $k'\!+1$}
		\Else
		\State $l \leftarrow l_{min}$
		\Do \Comment{\textit{Try to increase the window length by one}}
		\State $k \leftarrow k+1$
		\State $l \leftarrow l+1$
		\State Acquire $D(k)$
		\doWhile {${\mathcal D}({\mathcal J}_{k,l})$ satisfies CS 1-3 {\bf and} $k \leq \overline k_d$}
		\State Compute updated parameter estimate $\hat\mu(k)$  via RLS using ${\mathcal D}({\mathcal J}_{k-1,l-1})$	
		\State $k'\leftarrow k+1$
		\EndIf
		
		\EndWhile
		
	\end{algorithmic} 	
\end{algorithm}

Concerning the selection of the initial parameter guess $\hat\mu(0)$, the following observations are in order.
\begin{itemize}
	\item As previously stated, a good guess for the main power/irradiance gain $\mu_1$ is represented by $\hat\mu_1(0)=P_{nom}/1000$, where $P_{nom}$ denotes the nominal plant power \cite{pepe2017model,BIB:CDC13}. As pointed out in Remark \ref{rem:params}, it is appropriate to start with an underestimate of this value, e.g., $75\%$, to ensure faster parameter adaptation.
	\item As for the initial values $\hat\mu_2(0)$ and $\hat\mu_3(0)$, it is convenient to choose them so that $\mu_2(0)/\mu_1(0)$ and $\mu_3(0)/\mu_1(0)$ are equal to the central values of the intervals ${\mathcal S}_2$ and  ${\mathcal S}_3$ in \eqref{EQ:AtoBC}, respectively \cite{pepe2017model}.
\end{itemize}

\section{Forecasting}\label{sec:forecasting} 
In this section, we briefly describe how the estimated PVUSA model can be used in order to provide the generation forecasts used in the experimental part of this work. Let $k$ be a generic time instant in which a forecast is supposed to be computed and submitted, e.g., to the DSO. For a given time instant $j\geq k$, let $\hat W(j|k)=\{\hat I(j|k),\hat T(j|k)\}$ denote the weather forecast (irradiance and temperature) relative to time $j$ available at time $k$, where the irradiance forecast is projected on the panel surface using a-priori information on the plant orientation, if available, or a guess thereof taken from guidelines such as those in \cite{BIB:ORIENTATION}. The prediction of generated power for time instant $j$, computed at time $k$ using the parameter vector estimate $\hat\mu(q)$ available at time $q\leq k$, is given by
\begin{equation}\label{EQ:PHAT}
\hat P(j|k;q)=
\phi'\left(\hat I(j|k),\hat T(j|k)\right)\cdot\hat\mu(q) .
\end{equation}
In the following section, the forecasting performance of the PVUSA model estimated using the procedure detailed in Section \ref{sec:estimation} will be evaluated on the widely used Day-Ahead (DA) and Hour-Ahead (HA) forecasts \cite{BIB:STATEOFART}. The DA forecast is usually submitted at 6 am on the day before each operating day, which begins at midnight on the day of submission, and covers all
24 hours of that operating day. The HA forecast is usually submitted 105 minutes prior to each operating hour and provides an advisory forecast for the 7 hours of light (or the remaining ones, if less) of the same day after the operating hour. The time series representing the DA and HA forecasts can be constructed from the pointwise forecast \eqref{EQ:PHAT} by letting $j$ and $q$ vary in suitable sets. The details are omitted here for the sake of brevity and the reader is referred to Section 5 of \cite{pepe2017model}.

\section{Performance evaluation}
\label{SEC:PEVA}
In this section we introduce the performance assessment indices that will be used to evaluate the efficacy of the proposed method in the forecasting problems sketched in the previous section.

\subsection{Error measures}
For the sake of simplicity, a generic definition of the performance indices that will be used is given here. Details on how such indices are computed using a predictor such as \eqref{EQ:PHAT} in the specific contexts of DA or HA forecasting are provided in \cite{pepe2017model}.
%
Let $\hat P(j)$ represent the forecasted power and $P^m(j)$ the corresponding measured value. The following standard error measures are considered:

{\small \begin{align*} 
	&	RMSE  = \sqrt{\frac{1}{K}\sum_{j\in{\mathcal K}} \left( P^m(j) - \hat{P}(j) \right)^2 } \\
	&	MBE   = \frac{1}{K} \sum_{j\in{\mathcal K}} \left( P^m(j) - \hat{P}(j) \right)\\
	&	MAPE  = \frac{1}{K} \sum_{j\in{\mathcal K}}\abs*{\frac{P^m(j) - \hat{P}(j)}{P^m(j)}} \cdot 100\\
	&	NRMSE  = \sqrt{\frac{\sum_{j\in{\mathcal K}} \left( P^m(j) - \hat{P}(j) \right)^2 }{\sum_{j\in{\mathcal K}} \left( P^m(j) - \bar{P} \right)^2 }} \\
	&	R^2   = 1-NRMSE^2 \\
	&	RMSE_{NP}  = \frac{RMSE}{P_{nom}} \\
	&	MAPE_{NP}  = \frac{1}{K} \sum_{j\in{\mathcal K}}\abs*{\frac{P^m(j) - \hat{P}(j)}{P_{nom}}} \cdot 100 . 
	\end{align*}	 }\\
\noindent where ${\mathcal K}=\{1,\dots,K\}$ denotes the time span of the data set and $\bar{P}$ is the sample mean of the measured power.
The last two indices, i.e., $RMSE_{NP}$ and $MAPE_{NP}$, are normalized with respect to the nominal plant power $P_{nom}$ and are of practical interest for network operation. In particular, values lower than $10\%$ are considered  acceptable for network operation~\cite{BIB:WIDISS2014,coimbraoverview}.

\subsection{Benchmarks}
As an additional evaluation tool, the performance indices achieved using the proposed approach will be compared to those obtained using:
\begin{itemize}
	\item ODNP: the One-Day-ahead Naive Predictor\mbox{, i.e.,}
	\begin{equation}
	\hat P(j)=\hat{P}^{ODNP}(j) = P^m_{d-1}(j),
	\end{equation}
	where $P^m_{d-1}(j)$ denotes the measure of generated power recorded during the day before at the same time of day,
	\item SRLS: a PVUSA model estimated via a standard RLS algorithm in the complete information case, i.e., using actual measurements of generated power, irradiance and temperature.
\end{itemize}

\input{experiment}

\section{Conclusions}\label{sec:CONC}
In this paper, an efficient technique for estimating a forecasting model of photovoltaic power generation from limited information has been proposed. The approach is based on a set of tests performed on power data combined with a recursive estimation framework. It only exploits the time series of generated power and forecasts of temperature, the latter obtained from a meteorological service. The procedure especially fits the typical scenario where the network operator has no access to on-site measurements of irradiance and temperature, due to the large number of plants connected to the grid.

The algorithm has been extensively validated on two plants located in Italy, both on measured data and on forecasts of weather variables. The latter case reproduces a typical DSO scenario. Experiments worked out show very good forecasting performance, with limited computational burden.

Ongoing work addresses the aggregation of several plants covering large geographic areas. Due to a better quality of weather forecasts in this case, a significant accuracy improvement is expected. The integration of PV power generation forecasting in smart buildings and in microgrids will also be considered.

\section*{References}
\bibliographystyle{elsarticle-num}
\bibliography{../../bibliography/biblio}

\end{document}

%% file: experiment.tex
\section{Experimental results}
\label{SEC:EXP_RES}
In this section an extensive validation of the proposed procedure is presented. Two experiments have been run to evaluate the performance of the algorithm. In the first one, both model estimation and validation have been conducted using measured data (power and temperature for estimation, irradiance and temperature for forecasting) in order to assess the performance of the estimation procedure net of errors due to inaccuracies of weather forecasts. In the second, meteorological predictions have been used both for model parameter fitting and generation forecasting. The latter scenario corresponds to a typical DSO use case.

\subsection{Experiment set up}
For the two experiments performed, the following data sets have been used, respectively:
\begin{itemize}
\item[D1:] data from a PV plant P1 with nominal power $P_{nom} = \SI{960}{\kilo\watt}$p located in the campus of the University of Salento, in Monteroni di Lecce, Puglia, Italy (see \cite{BIB:MALVONI2016DATA} for details). Data, ranging from March 5th, 2012 to December 31st, 2013, consist of hourly samples of averaged measured power, air temperature and normal irradiance (the latter used only for comparison in the SRLS benchmark) ;
\item[D2:] data from a PV plant P2 with nominal power $P_{nom} = \SI{920}{\kilo\watt}$p located in Sardinia. Data, ranging from February 2nd, 2012, to May 1st, 2012, consist of hourly samples of averaged measured power, one day-ahead forecasts of air temperature and one day-ahead forecasts of normal irradiance. Information about the quality of such forecasts is reported in Table \ref{tab:weatherquality}.
\end{itemize}
\begin{table}[!htb]
	\centering
   \[	
	\begin{array}{cccccc}
	\toprule
	        & RMSE & MAPE & MBE & R^2 & NRMSE \\
	\midrule	
	\hat{I} & \SI{148}{\watt/\metre^2} & 77\% & \SI{29.1}{\watt/\metre^2} & 0.808 & 0.438 \\
	\hat{T} & \SI{1.9}{\degreeCelsius} & 23\% & \SI{0.9}{\degreeCelsius}  & 0.849 & 0.389 \\
	\bottomrule
	\end{array}	  	\]                                                                                                                                                                                                                                                                                                                                                                                                                                                                                                                                                                                                                                                                                                                                                                                                                                                                                                                                                                                                                                                                                                                                                                                                                                                                                                                                                                                                                                                                                                                                                                                                                                                                                                                                                                                                                                                                                                                                                                                                                                                                                                                                                                                                                                                                                                                                                                                                                                                                                                                                                                                                                                                                                                                                                                                                                                                                                                                                                                                                                                                                                                                                                                                                                                                                                                                                                                                                                                                                                                                                                                                                                                                                                                                                                                                                                                                                                                                                                                                                                                                                                                                                                                                                                                                  
	\caption{Quality indices of irradiance and air temperature forecasts for data set D2.}\label{tab:weatherquality}	
\end{table}
Therefore the data sets used for model estimation in the two cases are given by:
\begin{align*}
\mathcal{D}_1 &= \Big\{ \{ P^m(k),T^m(k),I^{cs}(k) \}, \; \in {\mathcal K}_1 \Big\}, \\
\mathcal{D}_2 &= \left\{ \{ P^m(k),\hat{T}(k),I^{cs}(k) \}, \; \in {\mathcal K}_2 \right\},
\end{align*}                                                                     where the sets of time indices $\mathcal{K}_1$ and $\mathcal{K}_2$ span the entire periods reported above for D1 and D2, respectively, with a sampling time $\tau_{s} = \SI{1}{\hour}$, and $I^{cs}(k)$ is generated using \eqref{EQ:ELIODON},\eqref{EQ:ICS}. Clearly, only time indices $k$ corresponding to hours of light were considered.

The initial parameter vector has been chosen according to the criteria in Section \ref{sec:estimation}, i.e., $\hat\mu_1(0) = 0.75~P_{nom}/1000$,  $\hat\mu_2(0) = \num{-1.34e-4}\cdot \hat\mu_1(0)$, and $\hat\mu_3(0) = \num{-3.25e-3}\cdot \hat\mu_1(0)$. Concerning the panel orientation angles $(\zeta,\psi)$ used in \eqref{EQ:ICS}, they have been chosen using a-priori knowledge: measurements of the panel angles and location, for P1 and P2, respectively. In particular, plant P1 is actually composed of two arrays with different orientations; for this plant an equivalent orientation has been estimated by averaging the respective angles, considering the nominal powers as weights. The parameters just described are summarized in Table \ref{tab:initparam}.

\begin{table}[!htb]
	\centering
	\[
	\begin{array}{c|ccccc}
	\toprule
				 & \hat{\mu}_1(0) & \hat{\mu}_2(0) & \hat{\mu}_3(0) & \psi & \zeta \\
	\midrule
	\text{P1} & 0.72 	& \num{-9.68e-5} & \num{-2.34e-3} &  \SI{10.6}{\degree} & \SI{-10}{\degree}\\
	\text{P2} & 0.690 & \num{-9.28e-5} & \num{-2.24e-3} &  \SI{27}{\degree}   & \SI{12.5}{\degree}\\ 
	\bottomrule
	\end{array}	
	\]
	\caption{Initial parameters and panel orientation angles.}\label{tab:initparam}
\end{table}

Concerning the choice of $\epsilon$, it is worth recalling (see Remark \ref{rem:params}) that  in order for CS test 3 to reject uniformly cloudy data with a CCF $\beta \leq \beta_0$, $\epsilon$ can be chosen approximately as
\begin{equation}
	\label{eq:epsilon_k}
	\epsilon = 1 - \frac{ P_{nom} }{1000} \cdot \frac{1}{\hat\mu_1} \cdot \beta_0,
\end{equation}
where $\hat\mu_1$ represents the currently available estimate of $\mu_1$. Therefore, we find it convenient to fix the CCF bound $\beta_0$ and adjust $\epsilon$ dynamically via \eqref{eq:epsilon_k} as soon as a new estimate $\hat\mu_1$ is computed.
In this respect, we observe that the range of variability of the CCF depends on the climate of the macro-area where the plant is located, which is usually available. For the Italian case, typical values of the CCF range from $0.5$ to $1$ \cite{BIB:SPENA}. In the experiments of this section, we choose $\beta_0=0.9$. However, higher/smaller values of $\beta_0$ within the typical variability range make the CS detection algorithm more/less selective. Therefore, an evaluation of this effect is also in order.

\subsection{Validation on measured data (D1)}
\label{SEC:MES_EVALUATION}
The proposed method (denoted as CSD) has been evaluated with reference to day-ahead (DA) forecasts \cite{pepe2017model} by taking actual measurements of meteorological variables as the respective forecasts. The performance is compared with that of both the ODNP and the SRLS. Initialization data are summarized in Table \ref{tab:initCSDMAGO1}.
\begin{table}[!htb]
\[
\begin{array}{r|l}
	\toprule	
	\text{Data set ID} 	&  \text{D1} \\	
	\text{PVUSA}	 &	 \hat\mu(0) = \begin{bmatrix} 0.72,& \num{9.68e-5}, & \num{-2.34e-3}	\end{bmatrix}'	\\
	{\beta_0}	& 	0.9\\
	\bottomrule
\end{array}
\]
\caption{Algorithm parameters for validation on measured data (D1)}\label{tab:initCSDMAGO1} 
\end{table}
 
\begin{figure*}[!htb]
	\PlotLabel
	\centering
	\includegraphics[width=1\textwidth]{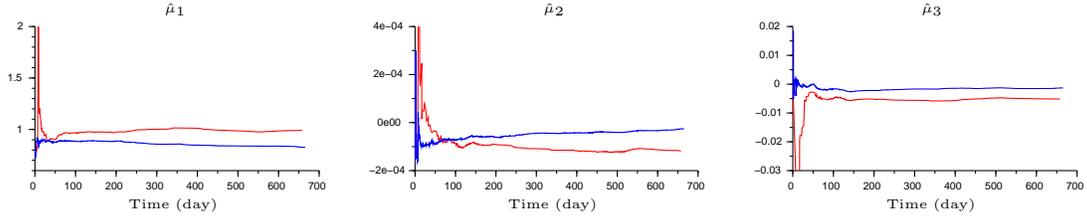}\\
	\caption{PVUSA parameters estimation using the CSD algorithm (red line) and a SRLS algorithm (blue line).}
	\label{fig:mes_param}
\end{figure*}

The time evolution of the parameters estimated using CSD and SRLS algorithms are shown in Figure~\ref{fig:mes_param}. Since the two algorithms use different data, namely theoretical irradiance for CSD and measured irradiance for SRLS, it is not surprising that parameters tend to slightly different values.

\begin{figure*}[!htb]
	\PlotLabel
	\centering
	\includegraphics[width=1\textwidth]{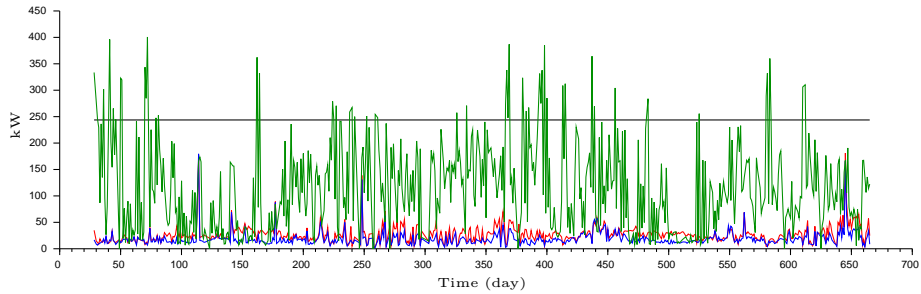} 
	\caption{(D1): $RMSE_d$ on DA~forecast. CSD (red), SRLS (blue) and ODNP (green). Black line represents the standard deviation of the measured power.}	
	\label{fig:mes_rmse}
\end{figure*}

As far as the forecasting performance is concerned, all error measures on DA predictions were computed over the period starting from day $28$, in order to guarantee at least a rough adaptation of the model parameters. In Table~\ref{tab:mes_errorcmp} the performance indices achieved by the proposed CSD approach are compared with SRLS and ODNP. Errors computed on CSD and SRLS are comparable and clearly show better performance with respect to the ODNP. In Figure~\ref{fig:mes_rmse}, the time evolution of the daily RMSE ($RMSE_d$) is, shown .

%
\begin{table}[tb]
 	\centering
	\begin{tabular}{llccc}
	\toprule
	\multicolumn{2}{c}{ \TwoRowCell{ Performance \\ Indices }}
										& CSD			& SRLS 		& ODNP		\\
	\midrule
	\parbox[t]{2mm}{\multirow{7}{*}{\rotatebox[origin=c]{90}{DA Forecast}}} %
	&$RMSE$ (\si{\kilo\watt})	& $31.0$		& $23.1$ 	& $143.2$ 	\\
	&$MAPE$							& $31\%$		& $26\%$ 	& $109\%$ \\
	&$MBE$ (\si{\kilo\watt})	& $-7.01$	& $-6.73$ 	& $3.00$		\\
	&$R^2$							& $0.98$		& $0.99$ 	& $0.65$	\\
	&$NRMSE$							& $0.13$  	& $0.10$ 	& $0.59$	\\
	&$RMSE_{NP}$					& $0.032$	& $0.024$ 	& $0.15$		\\
	&$MAPE_{NP}$					& $2.2\%$	& $1.5\%$ 	& $8.4\%$ 	\\	
	%
	\bottomrule
	\end{tabular}
	\caption{Performance comparison of CSD, SRLS and ODNP computed starting from day $28$ (D1).}
	\label{tab:mes_errorcmp}
\end{table}


\begin{figure*}[!htb]
	\PlotLabel
	\psfrag{8}{\hspace*{-3mm}\tiny{Day $8$}}
	\psfrag{9}{\hspace*{-3mm}\tiny{Day $9$}}
	\psfrag{10}{\hspace*{-3mm}\tiny{Day $10$}}
	\psfrag{32}{\hspace*{-3mm}\tiny{Day $32$}}
	\psfrag{91}{\hspace*{-4mm}\tiny{Day $91$}}
	\psfrag{419}{\hspace*{-4mm}\tiny{Day $419$}}
	\centering
	\includegraphics[width=1\textwidth]{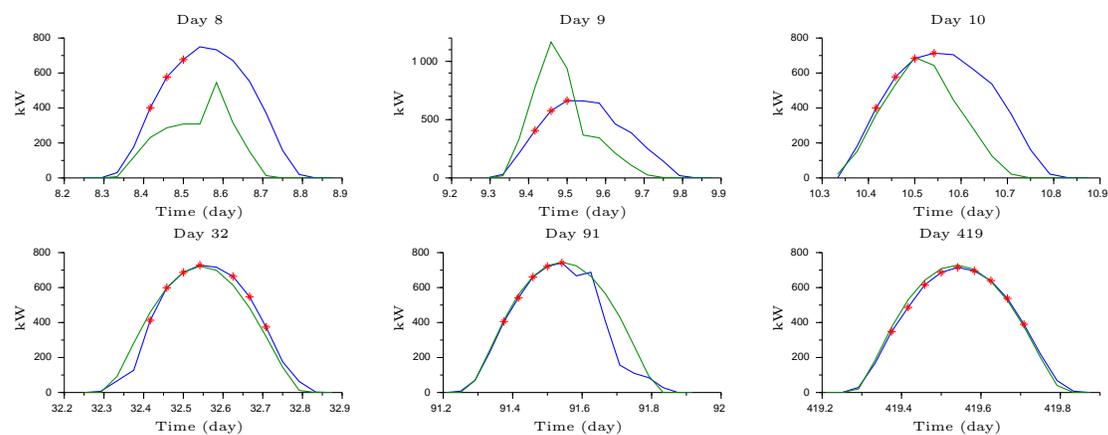} \\
	\caption{Visual representation of an algorithm run (D1). Measured power is in blue, current predicted clear-sky power is in green, red markers denote detected clear-sky windows.}
	\label{fig:detection}
\end{figure*}

A visual representation of the algorithm behavior with special attention to clear-sky detection is shown in Figure~\ref{fig:detection}. In those graphs, sequences of red markers denote time windows in which the measured power is detected as being generated under a clear-sky condition. The adaptation of model parameters is apparent from a comparison of the measured and predicted power in successive clear-sky periods. With reference to Figure~\ref{fig:detection}, in day $8$ the first clear-sky window is detected: note that $\hat{P}^{cs}$ is much lower then $P^m$. During day $9$ the second clear-sky window is detected, in this case the model overestimates the actual generated power. On day $10$ the model fit has largely improved. The remaining plots show other three, non consecutive days: days $32$ and $419$ are completely clear-sky; day $91$ is a partially clear-sky day, in which about a half of the samples is rejected by the algorithm.

\begin{figure*}[!htb]
	\PlotLabel
	\psfrag{ddd 11}{\hspace*{-2.5mm}\tiny{Day $32$}}
	\psfrag{ddd 222}{\hspace*{-2.5mm}\tiny{Day $140$}}
	\psfrag{ddd 333}{\hspace*{-2.5mm}\tiny{Day $277$}}
	\centering
	\includegraphics[width=1\textwidth]{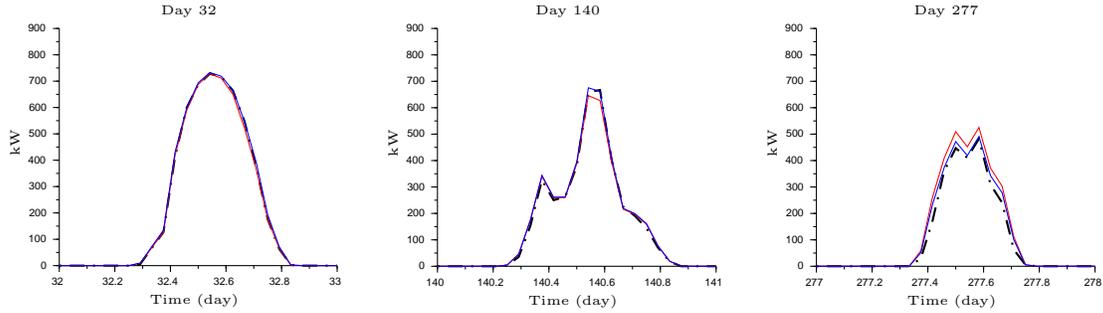} \\
	\caption{(D1): Comparison between the measured power (dash dot line), DA CSD forecast (red line) and DA SRLS forecast (blue line). From right to left, a clear-sky day, an overcast day and a partially clear-sky day.}
	\label{fig:mes_genpower}
\end{figure*}

Finally, in Figure \ref{fig:mes_genpower}, DA forecasts provided by CSD and SRLS during three different days and under three different weather conditions are compared with the measures of generated power. 

\subsection{Influence of $\beta_0$}
\label{sec:beta0}
To show the influence of the choice of $\beta_0$ on the behavior of the algorithm, the following experiment on data set D1 has been performed. The presence of a certain amount of power data generated under uniform cloudiness has been simulated by scaling down a given fraction of the power data collected during days which appeared to be clear-sky by inspection. The power curve related to each of such days has been scaled by a factor ranging from $0.5$ to $0.9$. Three different datasets have been generated, each containing a different fraction of scaled data, as described in Table~\ref{tab:learningset}. For each data set, the model estimation experiment has been repeated several times by varying $\beta_0$ from $0.40$ to $0.95$ with steps of $0.05$.
%
\begin{table}[!htb]
	\centering
	\begin{tabular}{c|ccc}
	\toprule
	Data set ID  & D1  	& D1$_1$  & D1$_2$ \\
	POD & $0\%$ & $5\%$ & $14\%$ \\
	\bottomrule	
	\end{tabular}
	\caption{Data sets used in the evaluation of the effect of $\beta_0$. POD denotes the percentage of scaled clear-sky data introduced.}
	\label{tab:learningset}
\end{table}
For the sake of fairness, original data from D1 have been used in all cases to compute forecasting errors. 


%
%
%
%

\begin{figure*}[!htb]
	\psfrag{nd}[][][.7][0]{}
	\psfrag{b0}[][][.7][0]{$\beta_0$}
	\centering
	\includegraphics[width=1\textwidth]{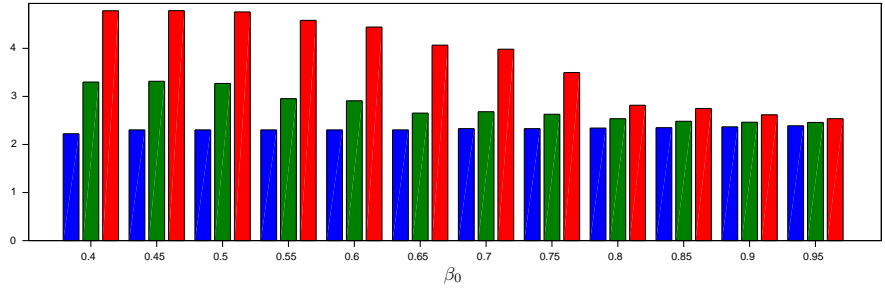} \\
	\caption{Percentage of data detected as clear-sky vs. $\beta_0$. Different data sets are depicted using different colors, D1 in blue, D1$_1$ in green and D1$_2$ in red.}
	\label{fig:csd_grid_data}
\end{figure*}

\begin{figure*}[!htb]
	\fragCSDGridError
	\centering
	\includegraphics[width=1\textwidth]{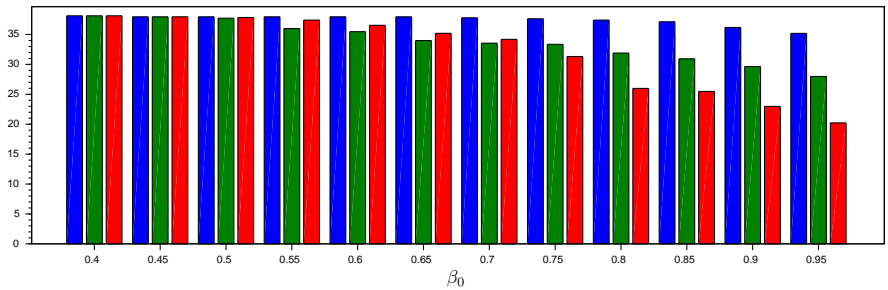} \\
	\caption{$\MAPEpp$ vs $\beta_0$. Different data sets are depicted using different colors, D1 in blue, D1$_1$ in green and D1$_2$ in red.}
	\label{fig:csd_grid_error}
\end{figure*}

Figure~\ref{fig:csd_grid_data}  shows the percentage of generated power measurements detected as clear-sky by the algorithm in the different data sets for varying $\beta_0$. Figure~\ref{fig:csd_grid_error} depicts the corresponding value of the $\MAPEpp$ on DA forecasts. When $\beta_0$ increases, the CSD algorithm becomes more selective. This fact is reflected in the $\MAPEpp$, which is lower in general for higher $\beta_0$. For given $\beta_0$, the error increases with the percentage of uniformly cloudy days. This phenomenon becomes less apparent as $\beta_0$ increases.


\begin{figure*}[!htb]
	\PlotLabel
	\centering
	\includegraphics[width=1\textwidth]{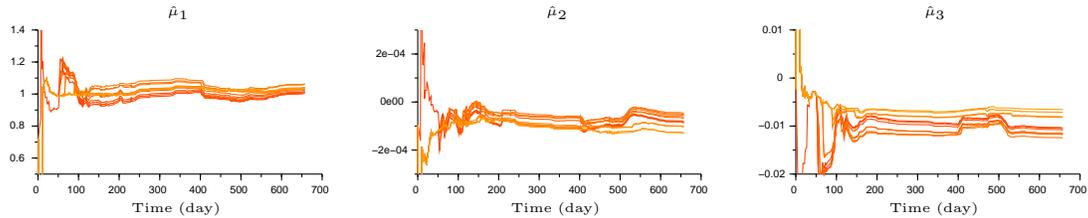} \\
	\caption{Parameter estimates vs. iteration for different choices of $\beta_0$. Values of $\beta_0$ are depicted using a color map which ranges from red to yellow, denoting the minimum and the maximum, respectively. Model identification is performed using D1$_2$.}
	\label{fig:csd_grid_param}
\end{figure*}

In Figure~\ref{fig:csd_grid_param}, the evolution of the parameter estimates performed on D1$_2$ for varying $\beta_0$ is reported. Parameter estimates tend to become almost stationary in all cases. Mean values and variances of $\hat{\mu}$ are reported in Table \ref{tab:meanstd}. 
Notice that $\hat{\mu}_1$ shows lower sensitivity to $\beta_0$ compared to $\hat{\mu}_2$ and $\hat{\mu}_3$.
\begin{table*}[!htb]
	\centering
	\[	
	\begin{array}{ccc}
	\toprule
	\beta_0 & \text{Mean of } \hat{\mu} & \text{Standard deviation of } {\hat{\mu}} \\
	\midrule	
	0.40	  & \begin{bmatrix} 0.970 & \num{-8.07e-5} & \num{-9.587e-3} \end{bmatrix}' & \begin{bmatrix} \num{4.82e-2} & \num{2.55e-5} & \num{3.14e-3} \end{bmatrix}' \\
	0.65	  & \begin{bmatrix} 1.026 & \num{-5.96e-5} & \num{-1.078e-2} \end{bmatrix}' & \begin{bmatrix} \num{2.82e-2} & \num{2.67e-5} & \num{2.09e-3} \end{bmatrix}' \\
	0.90	  & \begin{bmatrix} 1.015 & \num{-1.07e-4} & \num{-6.533e-3} \end{bmatrix}' & \begin{bmatrix} \num{1.66e-3} & \num{2.06e-5} & \num{7.83e-4} \end{bmatrix}' \\
	\bottomrule
	\end{array}
	\]
	\caption{Mean values and standard deviations of the parameters estimated using D$1_2$ and three different values of $\beta_0$ (computed from day $28$).}
	\label{tab:meanstd}
\end{table*}

\begin{figure*}[!htb]
	\PlotLabel
	\psfrag{day 55}{\hspace*{-3mm}\tiny{Day $55$}}
	\psfrag{day 90}{\hspace*{-3mm}\tiny{Day $90$}}
	\psfrag{day 99}{\hspace*{-3mm}\tiny{Day $99$}}
	\centering
	\includegraphics[width=1\textwidth]{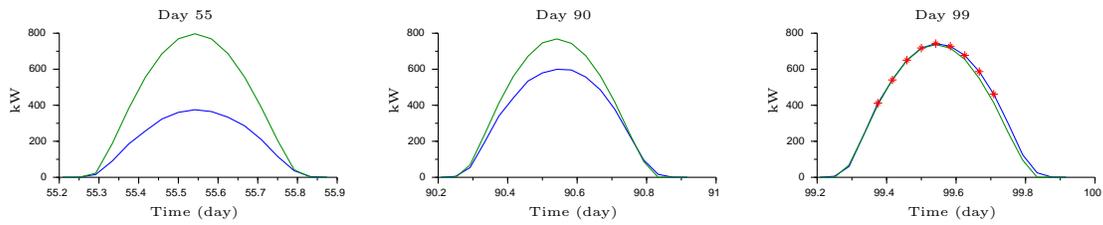} \\
	\caption{Visual representation of the role of CS test~3 with $\beta_0 = 0.9$ using data in D$1_2$. Plots show three different days in which tests~1 and 2 are satisfied. Measured power is in blue, theoretical clear-sky generated power is in green, red markers denote detected clear-sky windows.}
	\label{fig:cs3detection}
\end{figure*}

Figure~\ref{fig:cs3detection} depicts measured power and predicted clear-sky power during three uniformly cloudy days belonging to D$1_2$ for $\beta_0=0.9$. All data satisfy CS tests 1 and 2. CS test 3 is satisfied only for the last day, in which power data was obtained using a scaling factor greater or equal to 0.9. 


\subsection{Validation on predicted data (D2)}
\label{sec:dsoscenario}
In this section a typical DSO scenario is reproduced, in which it is assumed that measurements of weather variables are not available at the plant site. Therefore, measured power and temperature forecasts are used to estimate the plant parameters, while irradiance and temperature forecasts are used to cast predictions of generated power. The algorithm configuration parameters are reported in Table \ref{tab:initCSDMAGO2}.

\begin{table}[!htb]
\[
\begin{array}{r|l}
	\toprule	
	\text{Data set ID} 	& \text{D2} \\	
	\text{PVUSA}	 	& \mu(0) = \begin{bmatrix} 0.69,& \num{9.28e-5}, & \num{-2.24e-3}	\end{bmatrix}'	\\
	{\beta_0}	&  0.9\\
	\bottomrule
\end{array}
\]
\caption{Algorithm parameters}\label{tab:initCSDMAGO2} 
\end{table}

\begin{figure*}[tb]
	\PlotLabel
	\centering
	\subfloat[][\emph{DA forecast.}]
	{\includegraphics[width=1\textwidth]{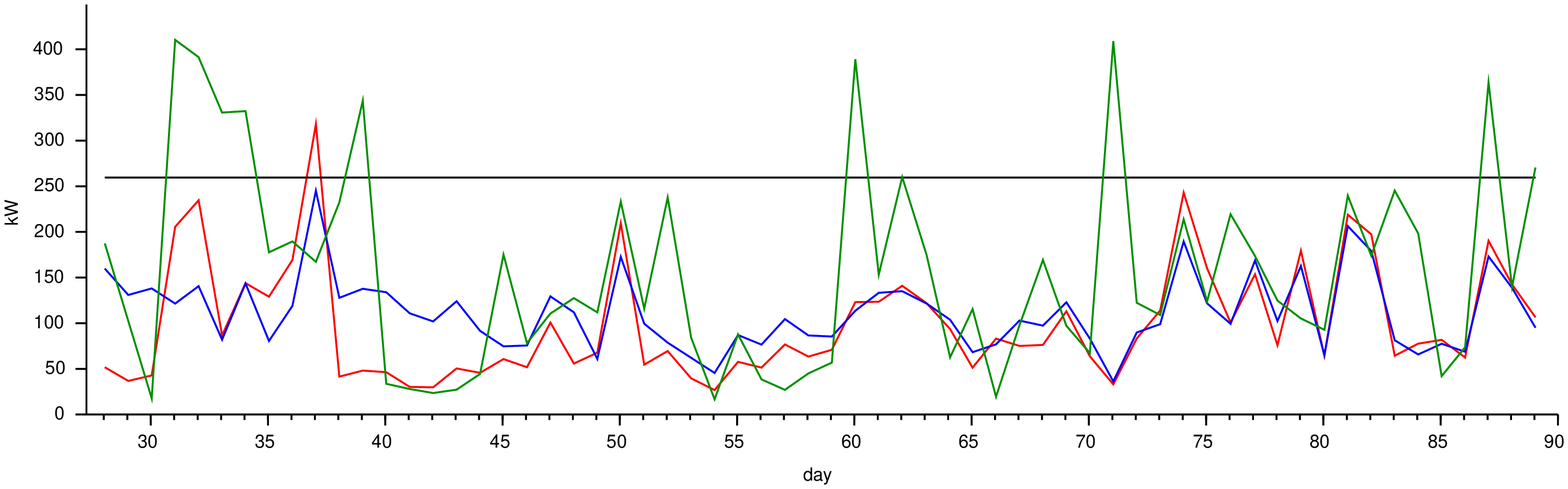} \label{fig:dso_rmse_da} } \\
	\subfloat[][\emph{HA forecast.}]
	{\includegraphics[width=1\textwidth]{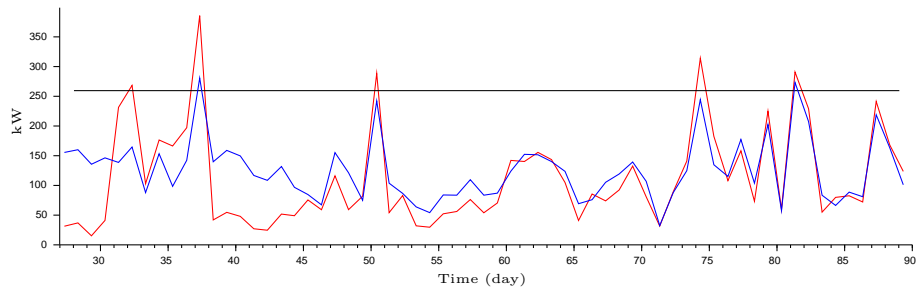} \label{fig:dso_rmse_ha} } 
	\caption{$RMSE_d$ on DA~forecast (top figure) and HA~forecast (bottom figure). CSD algorithm is in red, SRLS in blue and ODNP in green. The black line represents the standard deviation of the measured power.}	
	\label{fig:dso_mes_rmse}
\end{figure*}

In this scenario, the performance of the proposed method has been evaluated with reference to both day-ahead (DA) and hour-ahead (HA) forecasts, and compared with the performance achieved by SRLS and ODNP.
Forecasting error measures are reported in Table \ref{tab:dso_errorcmp} and Figure \ref{fig:dso_mes_rmse}. While ODNP still has the worst performance indices, CSD performs even better then SRLS. However, it should be observed that forecasting errors in this case are to a large extent due to the quality of weather reports (see Table \ref{tab:weatherquality}).

\begin{table}[tb]
 	\centering
	\begin{tabular}{llccc}
	\toprule
	\multicolumn{2}{c}{ \TwoRowCell{ Performance \\ Indices }}
										& CSD		& SRLS & ODNP		\\
	\midrule
	\parbox[t]{2mm}{\multirow{7}{*}{\rotatebox[origin=c]{90}{DA Forecast}}} %
	&$RMSE$ (\si{\kilo\watt})	& $117.9$	& $118.5$ 	& $193.3$ 	\\
	&$MAPE$							& $58.8\%$	& $55.2\%$  & $85.6\%$ 	\\
	&$MBE$ (\si{\kilo\watt})	& $-7.69$	& $35.6$ 	& $-5.6$		\\
	&$R^2$							& $0.799$	& $0.797$ 	& $0.458$	\\
	&$NRMSE$							& $0.448$   & $0.451$ 	& $0.736$	\\
	&$RMSE_{NP}$					& $0.128$	& $0.129$ 	& $0.201$	\\
	&$MAPE_{NP}$					& $8.3\%$	& $9.8\%$ 	& $12.4\%$ 	\\	
	\midrule
	\parbox[t]{2mm}{\multirow{7}{*}{\rotatebox[origin=c]{90}{HA Forecast}}} %
	&$RMSE$ (\si{\kilo\watt})	& $138.2$	& $136.2$ 	& -			\\
	&$MAPE$							& $52.1\%$	& $46.0\%$ 	& -			\\
	&$MBE$ (\si{\kilo\watt})	& $-25.8$	& $33.0$ 	& - 			\\
	&$R^2$							& $0.655$	& $0.665$ 	& -			\\
	&$NRMSE$							& $0.588$	& $0.579$ 	& -			\\
	&$RMSE_{NP}$					& $0.150$	& $0.148$ 	& -			\\
	&$MAPE_{NP}$					& $10.0\%$	& $11.9\%$ 	& -		 	\\
	\bottomrule
	\end{tabular}
	\caption{Performance comparison of CSD, SRLS and ODNP computed starting from day $28$ (D2).}
	\label{tab:dso_errorcmp}
\end{table}

\begin{figure*}[tb]
	\PlotLabel
	\psfrag{day 42}{\hspace*{-3mm}\tiny{Day $42$}}
	\psfrag{day 52}{\hspace*{-3mm}\tiny{Day $52$}}
	\psfrag{day 71}{\hspace*{-4mm}\tiny{Day $71$}}	
	\centering
	\includegraphics[width=1\textwidth]{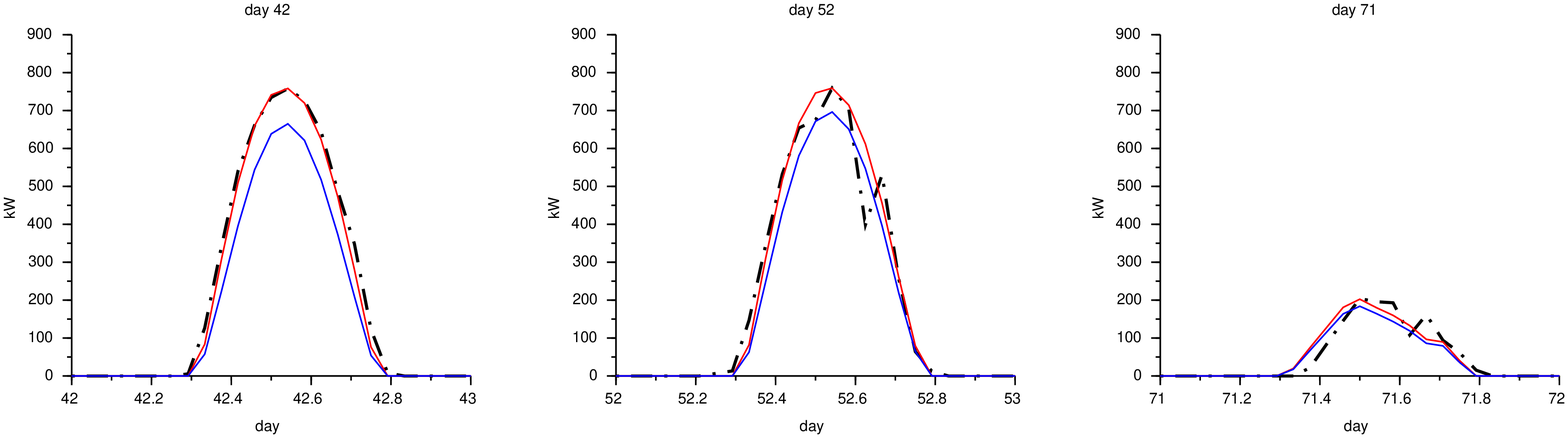} \\
	\caption{(D2): Comparison between measured power (dash dot line), DA CSD forecast (red line) and DA SRLS forecast (blue line). From right to left, a clear-sky day, an overcast day and a uniformly overcast day.}
	\label{fig:dso_genpower}
\end{figure*}

Three examples of DA forecast computed using CSD approach and SRLS during different weather conditions are shown in Figure \ref{fig:dso_genpower}.

\subsection{Further remarks}
\label{sec:discussion}
With reference to Tables~\ref{tab:mes_errorcmp} and~\ref{tab:dso_errorcmp}, it is important to observe that the normalized errors ($MAPE_{NP})$ computed on DA forecasts are below $10\%$, which demonstrates viability for network operation. Furthermore, the performance indices achieved by CSD are very close to those obtained by SRLS, i.e., via a PVUSA model estimated using measured irradiance.

Concerning the role of the tunable parameter $\beta_0$, results in Section~\ref{sec:beta0} show that the estimate of the main power/irradiance gain $\mu_1$ is quite robust with respect to $\beta_0$, and moreover the parameter estimates $\hat\mu$ tend to converge regardless of the value of $\beta_0$. Even for small $\beta_0$, i.e., when the algorithm is not selective and CS test~3 is satisfied even for heavy uniform cloudiness, CSD is able to provide reasonably accurate forecasts. Increasing the values of $\beta_0$, the algorithm tends to reject more and more data measured under a uniformly cloudy sky, resulting in an improvement of the forecast quality.

The proposed algorithm has been implemented in Scilab \cite{bib:scilab}. Each iteration took on average less than one second on an i7 2.6 Ghz processor, thus demonstrating that the approach carries an extremely low computational burden.